# Radiation Effects on Stepper Motors and Lubricants: A Review

Logan Jackson[1], Victor Boyer[2], Tanner Rima[3], Edward Cazalas[1, *]

1. Nuclear Engineering Program, Department of Civil and Environmental Engineering, 110 Central Campus Dr. STE. 2000, Salt Lake City, University of Utah, 84112.
2. Department of Mechanical Engineering, 1495 E 100 S, Salt Lake City, University of Utah, 84112.
3. Department of Atmospheric Sciences, 135 S 1460 E, Rm 819, Salt Lake City, University of Utah, 84112.

* Corresponding Author, Email: edward.cazalas@utah.edu

**Abstract**

*The use of electrical motors and other remote systems are important tools in radiation environments. Certain harsh radiation environments, such as particle accelerators, require the use of remote systems during operation. Stepper motors are one motor, in particular, that have acquired interest in the nuclear field for use in these remote systems. The stepper motor's precision in rotation and holding is a highly desired quality. In this review, the stepper motor is examined for use in radiation environments. The different mechanical parts are introduced and examined from the broader literature of previous work. We attempt to explain the different radiation destructive mechanisms involved with each part of the stepper motor and how the mechanics are affected. Current and past research is explored, identifying either radiation hard alternatives or thresholds of the materials involved, and a general review of stepper motor radiation effects as available in literature.*

# 1. Introduction

Remote systems are often an important part to work within harsh radiation environments. High radiation fields and high energy particles make it dangerous for direct human intervention in these areas, including in decay radiation fields. Activation via neutron reaction, especially in high energy beamlines, can cause longer lived radiation hazards. Several methods have been developed to minimize dose or prevent exposure all together. These methods have evolved from the primitive 'reach rod', to the glove box, to fully remote controllers like the 'mechanical master/slave manipulator', and other standard electromechanical devices. [1] The use of these systems, their components, and their radiation tolerance has been a large consideration of research and safety of particular radiation work. [2]

The stepper motor is an electromechanical device that has been a workhorse, particularly in remote systems. Stepper motors are used in the collimation system of the Large Hadron Collider (LHC) at CERN. The stepper motor excels at high precision and accuracy in movement and positioning. The motor allows for rotation in discreate steps and therefore does not need position feedback in order to maintain a high level of accuracy. These characteristics make it valuable for use in specific radiation environments, including space applications. [3] [4] In general, the stepper motor is quite radiation resistant when excluding the controller systems. Only specific parts have been shown to have significant degradation. These parts include the ball bearing lubricant, the wire coating, and the permanent magnets. [5][6] Many of these parts already have a limited lifetime due to the normal wear and tear of operation. But this lifetime can be shortened through radiation damage. Throughout this review the individual parts and the stepper motor itself will be examined and evaluated through the work of many different authors.

# 2. Mechanisms of Damage

To understand the effects of ionizing radiation on stepper motors, first the damage mechanics must be explored. The effects of radiation damage in a stepper motor can take many forms. In general, the degradation of materials creates malfunctions within specific systems of the motor so therefore, an understanding of radiation damage and radiation interactions must first be understood. There are many different mechanisms and interactions that ionizing radiation has with matter. This is an extensive study, and many textbooks have been written about radiation interactions. Of interest, is how radiation interacts with the target material and what mechanisms are involved in breakdown and change of these materials.

The most fundamental mechanism of radiation damage in materials is Threshold Displacement Energy (TDE) or Direct Displacement Damage (DDD). This is often also referred to as just the 'displacement energy'. In the simplest terms, this is the minimum recoil energy required for an atom to leave its original location and leave behind a defect in the material. [7] This is typically done via heavy charged particles or neutron interaction. The displaced atom creates a deformity in the crystal structure and therefore weakens the material. This displacement can often result in annealing where the displaced atom finds its original location (or other dislocation site) via diffusion. The displacement can also result in further damage in the creation of large vacancy clusters or vacancy loops from a high rate of dislocations. In these cases, the damage has manifested into a larger chain of damage. [7]

**Polymers**

Polymers when introduced to radiation can undergo chain scission, causing breakage in polymer chains that can lead to more brittle, less durable plastics. Typically, such radiation induced weakening is irreversible but may require a great deal of radiation exposure to result in the damage, as most polymers are highly resistant to radiation. In a materials review published by Shulman and Ginell,[8] the effects of gamma radiation on certain polymers were found to support this generalization of high resistivity with some notable exceptions. Teflon, as used in stepper motors has a significantly lower tolerance to radiation, experiencing mild to moderate damage at doses around $2 - 4 \times 10^4$ rad and moderate to severe damage as low as doses greater than $5 \times 10^4$ rad. With Teflon being utilized in stepper motors, exposure to radiation dose may over time cause a degradation of the materials and lead to a loss of function. The study states that materials exposed to this level of dose should have limited use, as their functionality may decrease past the doses outlined as severe damage. [8]

Other polymers used in stepper motors such as polyimide and polyamides have significantly higher resistance to radiation damage. White et al performed a radiation robustness experiment on polyamide-imide polymers and found that at doses up to 100 Gy caused no significant decrease in the functionality of the polymer. [9] This level of resistance to radiation is common amongst different classes of polymers and the dose is rather low compared to other studies. In high radiation environments, better performance will be found using thermosetting or condensation polymers from the polyimide-amide groups such as Nylon, Vespel and Kapton.

**Metals**

Metals, when exposed to radiation, have many of the same interactions that will be discussed with magnets. As the material is exposed to radiation, particle collisions produce a thermal spike of displaced atoms that can transfer energy outwards causing internal damage. Nordlund et al reviewed the current knowledge on radiation damage to several materials and noted that when a metal is introduced to radiation, the thermal spike acts nearly like a thermodynamic system with a Maxwell-Boltzmann like energy distribution. [7] Much of the initial damage produced by radiation is mitigated by recrystallization, however some of the produced damage remains inside the metal. Small defects produced by these cascades can result in damage that can cause corrosion, embrittlement and acceleration of oxide formation within the metals. Additionally, metal atoms can undergo transmutation via neutron reaction. This may lead to changes in the chemical state of the atoms as it decays and lead to a less stable structure of the metal materials. [7]

## 3. Stepper Motor Materials and Construction

In understanding how the stepper motor is affected by radiation it is first important to understand how the motor works and the different parts involved. Stepper motors operate by splitting the rotation of the motor into individual steps. There are three different kinds of stepper motors available. Variable reluctance, permanent magnet, and hybrid. All three utilize a similar method of operation, but we will be focusing on hybrid stepper motors. Hybrid stepper motors follow a NEMA # sizing characterization, where the number designates the side length of the motor face multiplied by 10.

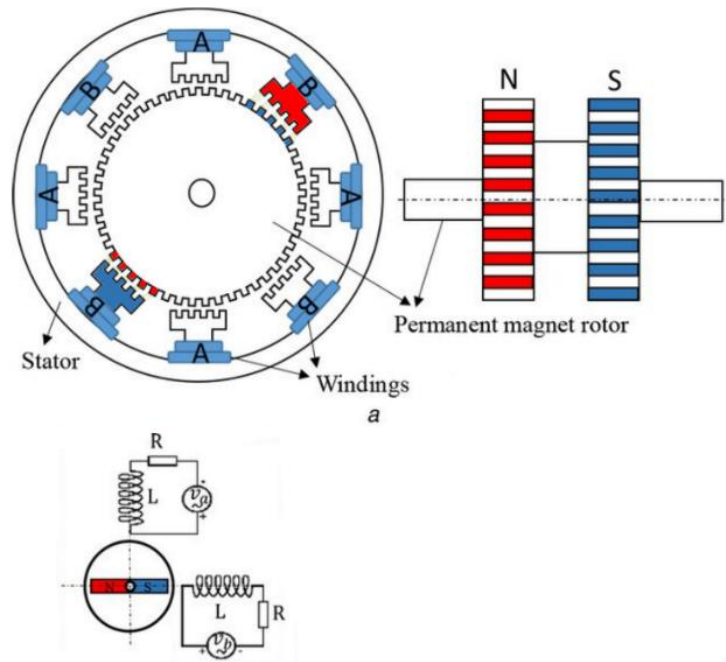

*Figure 1 - Image of stepper motor windings and permanent magnet. A and B indicate windings that are energized together. The rotor has a magnet of alternating North and South pole facing teeth, allowing for smaller step sizes within the motor. [8]*

An electric pulse is fed into the motor, energizing the coils and causing the rotor teeth to align with the stator teeth. This advances the rotor one step. A new pulse energizes the next set of stator teeth, and the rotor advances an additional step. [10] A visual portrayal of this can be seen in Figure 1. The most common step size is 1.8°. Due to this, one pulse gives 1.8° or rotation, 10 pulses rotates 18°, and 100 pulses rotates 180°. Continuous rotation is achieved by supplying the electric pulses at a specific frequency. Increasing the frequency results in faster rotation and decreasing the frequency results in slower rotation. Because of this, stepper motors offer very precise rotation control and high torque holding characteristics.

Stepper motors are constructed by placing a magnetic disk between two rotors on the rotor shaft. This rotor assembly is suspended by two bearings on the front and rear faces. Both front and rear faces are attached to the stator, which encircles the rotor assembly. The magnetic field produced by the copper windings on the stator is what makes the rotor turn. [11] Figure 2 shows an uncoiled schematic of the different stepper motor parts and their location in construction.

---

The ball bearings used in stepper motors are deep groove ball bearings, typically made up of 52100 chrome steel. Approximately 85% of ball bearings use a grease lubricant that follows these proportions: 85% mineral oil, 10% thickener, and 5% other additives. The thickener is frequently lithium stearate, but can also be calcium, aluminum, and polyurea based. The amount of grease contained within the bearing falls between 25% and 35% of the internal volume. Dimensions and approximate grease volumes are listed in inches and millimeters in **Error! Reference source not found.**.

Table 1 - Ball Bearing Dimension and Grease Volume

| Motor Size | ID | OD | Height | Grease Volume |
|---|---|---|---|---|
| 8 | 4mm | 16mm | 5mm | 0.235 mL |
| 11,14,17 | 5mm | 19mm | 6mm | 0.4-0.47 mL |
| 23 | 0.25 in | 0.62-0.75 in | 0.28 in | 0.376 mL |
| 34 | 0.5-0.6 in | 1.125 in | 0.312 in | 2.156 mL |
| 42 | 0.75 in | 1.625 in | 0.44 in | 10.65 mL |

The front and rear motor faces are typically made from aluminum. These support the rotor assembly. The shaft is made of mild steel, and the rotor and stator are made of silicone steel laminations (Fe: 19%, Ni: 78%, Si: 3%).

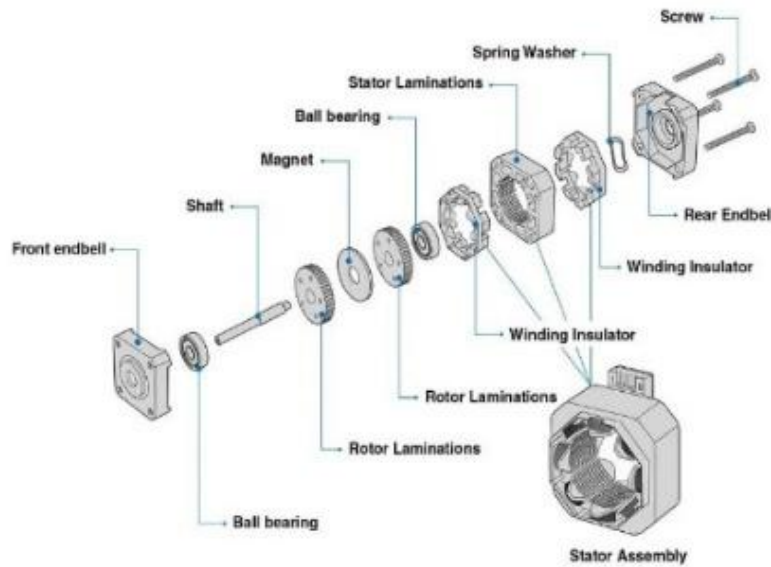

[2]

---

[2] Moons', "Step Motor – Basic Structure & Operation." Accessed: Jun. 23, 2025. [Online]. Available: https://www.moonsindustries.com/article/basic-structure-and-operating-principle-of-stepper-motor. Citation: [11]

*Figure 2 - Uncoiled image of stepper motor parts. [11]*

The stator windings are composed of copper wire (Cu: 99.3%, remainder: O) with an insulative coating approximately 0.0010 in. thick. The common kinds of insulation, organized by temperature ratings, are 105°C: formvar (vinyl-acetal film), 155°C: polyurethane, 200°C: polyamide imide, 240°C: polyimide, 260°C: Nylon/Teflon.

Typical Stepper Motor Component List
- Front and rear faces/end bells
- Stator
- Ball bearings
- Drive shaft
- Rotor cups/laminations
- Magnet
- Magnet wire
- Plastic magnet wire frame
- Spring washer
- Screws
- Lead wires
- Greasing

# 4. Radiation Sensitive Parts of Electrical Motors

Many of the materials of stepper motors are quite resistant to radiation. All materials will break down in some capacity, but mostly due to mechanical effort. There have been few studies on the radiation effects of whole electromagnetic motors in an attempt to classify operation thresholds for these motors. Discussion on the whole motor behavior is presented, but focus will be given to individual parts within the system and how these parts are affected in radiation environments. These radiation sensitive parts, likely to affect the overall lifetime of the motor, will be discussed throughout.

First it is important to mention that many of the electrical components included in the motor controller and driver are extremely sensitive to radiation and are only briefly discussed within this review. These components include semiconductor devices often seen in computer systems. Because of the particular sensitivity of most computer electronics, these systems are often placed far away from the radiation environment and connection is preserved through long wires. Further discussion of these systems and their possible difficulties resides in Section 5.

The first notable radiation sensitive part is the ball bearing lubricant. This being a typically organic material, it is very susceptible to degradation. Without proper lubrication characteristics, the motor bearings will experience increased friction when rotating. This causes wear on the bearing balls and ultimately, this wear could cause inconsistencies in rotation or vibrations affecting overall precision of the device. A possible alternative to standard stainless steel with lubrication bearing is a non-lubricated ceramic bearing. The alternative use of non-lubricated bearings is discussed in Section 9.

Wires are another notably susceptible part of the stepper motor. Wires are coated in insulating materials to improve their overall conducting/insulating properties by reducing the impact of interference from other close by wires. Insulation is a non-conductive material such as polyethylene or rubber. The insulation material of the wires can see significant degradation which can lead to interference and an increase in wire resistance. A more in-depth analysis is presented in Section 7.

Another notably sensitive part is the permanent magnet. Although there are not many studies on magnets taken straight from stepper motors, there are studies involving equivalent magnets of the same materials and strengths. Radiation exposure can cause decreases in the magnetic field of a magnet. Further explanation of the different magnets and how they behave is described in Section 6. This decrease can directly affect the mechanical output of the stepper motor in decreased torque and holding torque.

## 5. Radiation Effects on Motor Systems

Ionizing radiation primarily affects electric motors through two means: material degradation and electronic component interference. These effects occur from buildup of radiation dose over time and rarely have any immediate consequences. However, prolonged exposure and dose buildup can create performance decreases in stepper motors. The areas of greatest concern are the breakdown of electrical insulating coatings, weakening magnetic field strength, and breakdown of lubricants.
Previous studies have shown the effects of radiation on motor components, but extensive work has not been done on stepper motors specifically. Based on related works it is expected that a substantial decrease in magnetic field strength, breakdown of lubricant, and increase in wire resistance due to insulation breakdown will result in an overall decrease in performance. [12], [13], [14], [15], [16] These performance decreases would manifest in lower torque output and therefore RPMs in the motor.

These performance decreases have been seen in DC servo motors under gamma irradiations performed by Oak Ridge National Lab. [17] The DC servo motors were exposed during acceleration and deceleration cycles with radiation exposures of $2 - 3x10^9$ $R$ before failure. The failures were attributed to shortages in the motor windings. AC servo motors and stepper motors were also tested in this study. For both motors, no decrease in performance was observed up to $10^9$ $R$. The stepper motor in particular had been built and tested to $10^8$ $R$. After testing, an examination of the parts showed no serious degradation of individual parts. [17] Further and extended research is needed to determine the operation thresholds of stepper motors in radiation fields. Based on the performance degradation of the DC motor, a shortage in the wire windings might be the first indication of performance degradation in stepper motors.

Another system that must be considered with the uses of stepper motors in high radiation environments are the controls. The controls consist of controllers and drivers that are used to power and manipulate the motor. The electronics involved in the motor drive, controller, and power systems of the stepper motor are particularly sensitive to radiation which include most modern semiconductors devices. [18], [19], [20], [21] Because of this, these controls must either be shielded or removed from the environment entirely. Controller systems like the ones used with the collimators in the Large Hadron Collider at CERN prevent irradiation of the controllers and drivers by keeping them in radiation free areas and connecting the motors through long wires. Feedback positioning controls like position sensors must also be considered along with these other electronics. These position sensors provided verification of completed rotations. During normal operation or due to wear caused by it, the motor can experience malfunctions were the motor fails to complete full step rotations. When this happens, the motor and the controller will be misaligned, and the position of the motor will not be true to the controls. This presents its own set of challenges to verify complete accuracy of the motor in environments where a position sensor within the motor cannot be sustained. Several authors have done research to improve the positioning accuracy and controls of the stepper motors used in radiation environments using open loop systems or other methods. [22][23] Another method might involve the use of shielding to protect the whole motor, controls, and sensors. [24][25] It is important to protect these sensitive electronics when implementing stepper motors in radiation environments.

## 6. Radiation Effects on Permanent Magnets

**Neutrons**

Radiation effects on magnets have been the focus of numerous studies to ascertain the extent of demagnetization caused by the radiation. Previous studies have concluded that it is likely that radiation dose can cause thermal spikes within the materials that then lead to displacement cascades. These thermal spikes, typically of nanometer or larger scale radius will then allow for domain reversals leading to localized demagnetization. This effect is particularly noted in NdFeB permanent magnets where it displays a noticeable flux dependency. Cost et al. investigated the effects of fast neutrons on such magnets. When irradiated with a fluence of order of magnitude $\sim 1 x 10^{15}$ N/cm$^2$, it was found that after a full irradiation, magnets at 350K experienced nearly a 60% reduction in remanence and magnets at 426 K experienced an even greater decay at nearly a 95% loss in remanence post irradiation. [26] Significantly, the authors explained that when tested for potential annealing effects, for up to 4.25 hours there were no long-term effects in the remanence and that the effects of the irradiation are reversible. Their explanation of the temperature dependence on the degree of demagnetization was the proximity to the Curie temperature causing an increased likelihood of the displacement cascades to nucleate new reversal domains. [26]

It has been noted that past the threshold of fluences near $10^{16}$ neutrons/cm$^2$, NdFeB magnets receive permanent structural damage due to heat and annealing effects which disallow full recovery of magnetism. Cost et al. further determines that non-annealing decay in remanence allows for full recovery of magnetism, and due to pinning in the domain walls from the radiation process, would pin domain walls increasing the coercivity by up to 20% from the radiation. [26]

Klaffy and coworkers studied the effects of thermal neutrons on NdFeB magnets, hypothesizing a heightened decay in remanence due to the 3837 b (barns) absorption cross section of $^{10}B(n, \alpha)^7Li$, releasing an alpha particle that can induce thermal spikes within the magnet itself. [27] By incorporating the thermal spike model, they found that this thermal spike could approach or exceed the curie temperature of the magnet, reversing the domains. They concluded that the alpha particles emitted during the Boron capture of a thermal neutron played a significant role in the demagnetizing process.

**Gamma Radiation**

Gamma radiation may also cause damage within the magnet as well. Cao et al. exposed NdFeB and FeCrCo magnets to 1.35 MeV gamma rays from a $^{60}Co$ source with constant exposure until an accumulated dose of around 200 Mrad. [28] The results showed that gamma rays had little effect on the magnet, producing flux losses ranging from an increase of 0.5% to –2.5%. This was not the case for the FeCrCo magnets, which experienced degradation up to approximately 16.5%.

**Electrons**

Studies have been performed to determine the effects of electron radiation on magnets, but due to the nature of electrons producing secondary particles and radiation such as Bremsstrahlung have made it difficult to ascertain the exact nature of damage from electrons. Bizen at el. used NdFeB magnets that had been thermally stabilized to become more radiation resistant and exposed them to electron beam irradiation. They used beam currents of 4-5 mA, and 4 mA paired with 8 GeV and 6 GeV electron energies respectively. Their results indicated that the degradation of the magnet field was highly dependent on electron energy and localized around the irradiation center. They also concluded that due to the localized nature of electron induced radiation damage, there are effective methods to protect against the damage, listing methods such as replacing parts of the magnet with alloys or highly resistant magnets such as SmCo magnets. [29]

# 7. Radiation Effects on Electrical Conducting Wires

While most metals are radiation tolerant, the insulate coatings for wires are not as tolerant. In many studies the breakdown of the insulation polymers has been shown to occur with radiation dose. Embrittlement is a common side-effect of dose build up. [30] This may not lead to performance loss in stable environments, but under vibrational stress or, otherwise adverse conditions, could easily lead to the insulation coating breaking. In general, the insulation coatings are limited to their physical properties rather than their dielectric properties. [31] The embrittlement and removal of the coatings is the greatest cause for concern. In a worst-case scenario this could short the motor, or lead to performance loss. There is also the potential for induced currents within the system. While it is unlikely that this could cause a catastrophic failure, the induced current creates noise within the system leading to potential performance losses.[32]

Although the effect of radiation on the wire coatings is more prominent, under high radiation fields, the wire's electrical conductivity might experience effects. In a study looking at different copper alloys, the changes in electrical conductivity were determined under fast reactor neutron fields. The study showed both increases and decreases in electrical conductivity from different alloys. Most alloys saw small decreases in electrical conductivity besides the compositions that included nickel (Ni) which showed an increase in conductivity when exposed to radiation. [33] For the purposes of stepper motors, the insulation coating of the wire is a much greater concern as there are many wire connections within the motor system. Degradation and removal of the coating can result in electrical shorting, causing many problems in the motor operation. CERN has done research characterizing the thresholds of many different insulating wire materials. [34]

# 8. Radiation Effects on Lubricants

**Radiation Effects**

As detailed earlier, the bearings of the stepper motor are a vital part of the motor which allows the motor to rotate efficiently and with accuracy. Lubricants within the bearings are also vital to the smooth operation and rotation of these bearings. The lubricant is a particularly sensitive part of the motor often being made of organic materials. There are many properties that contribute to lubricant effectiveness. Important physical properties include density, viscosity, heat capacity, and thermal conductivity. Important chemical properties include solvency, dispersancy, detergency, antiwear, anticorrosion, frictional properties, and antioxidant capacity. [35] Some of these properties are determined by the base oils while some are determined by the additives.

As implied in the name, base oils are the base stock of lubricants. These base oils are usually petroleum or mineral oil derived. These base oils consist of hydrocarbons chains as seen in Figure 3. Though the base oil must possess the physical and chemical properties required for an effective lubricant, another and sometimes even more important part of the lubricant is the additive. The additive often overpowers the base oil in determining the lubricant's overall characteristics. There are many different additives, and the choice depends on the desired characteristics. [35] Though there are several studies on the radiation effects of lubricants, none point to specific materials or compounds within the lubricant that contribute to those specific effects. At the time of this report, most studies can only characterize the lubricant as a whole and correlations between similar lubricants are not ever reached.

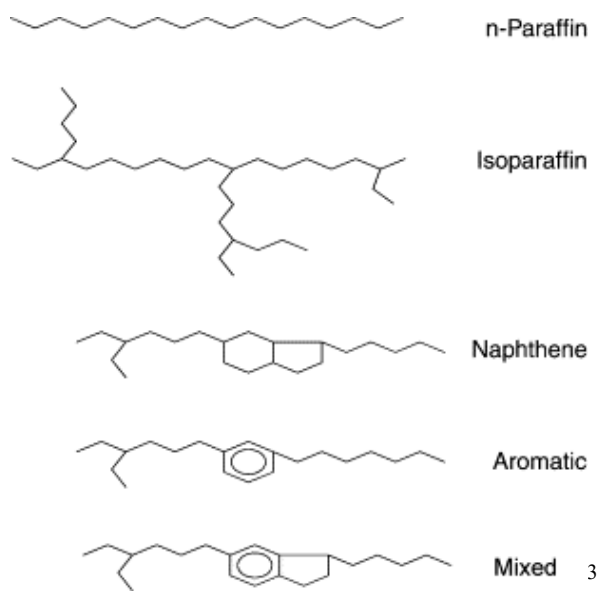

*Figure 3 - Molecular Structure of Base Oil Molecules.[35]*

---

[3] Reprinted from Tribology International, Vol 37 / Edition 7, Stephen M Hsu, 'Molecular basis of lubrication', 553-559. Copyright (2004), with permission from Elsevier.

It is often understood that the radiation resistance of a lubricant mostly depends on the base oil. Other studies conclude this is not true and believe that complex interactions between the base oils and the additives are a much more likely variable in radiation resistance. [36][37] Lubricants have many changes when exposed to radiation. These include color changes, gas production, evolution of acid products, changes in oil chain length, and thickener structure. Lubricants, when exposed to radiation, are found to have a change in consistency. This is a hardening or softening of the grease due to changes in the oil chain length or thickener structure. The relative consistency is described by the ratio of the irradiated sample and the unirradiated sample.

$$Crel(D) = \frac{C(D)}{C0}$$

Where C(D) is the consistency value (in mm/10) of the grease irradiated at a total dose D. $C0$ is the consistency value (in mm/10) of the unirradiated grease. [36]

Because each lubricant can only be characterized individually, it is best to look at studies or execute many experiments to get an idea of each lubricant's behavior under irradiation. One study looked at 9 grease samples tested in a mixed neutron and gamma-ray source in a TRIGA Mark II reactor. [36] Much like most studies in materials, each lubricant has a very specific reaction to radioactivity. Greases with similar base oils behave differently from each other indicating that it is difficult to draw conclusions on base oils alone. [38] Greases with manufacturer prescribed radiation thresholds often show extreme degradation before the threshold doses. [36] The greases that had manufacturer prescribed radiation thresholds were determined from gamma irradiation alone. Because of this, the characterization of these greases cannot be done based on dose alone. Specific particle interactions (alpha, beta, gamma, neutron) must be taken into account when determining radiation thresholds. Each lubricant will have different effects based on the specific particle. [36] Another study extrapolates these results to show the usefulness of each lubricant over a range of doses as seen in Figure 4. [37]

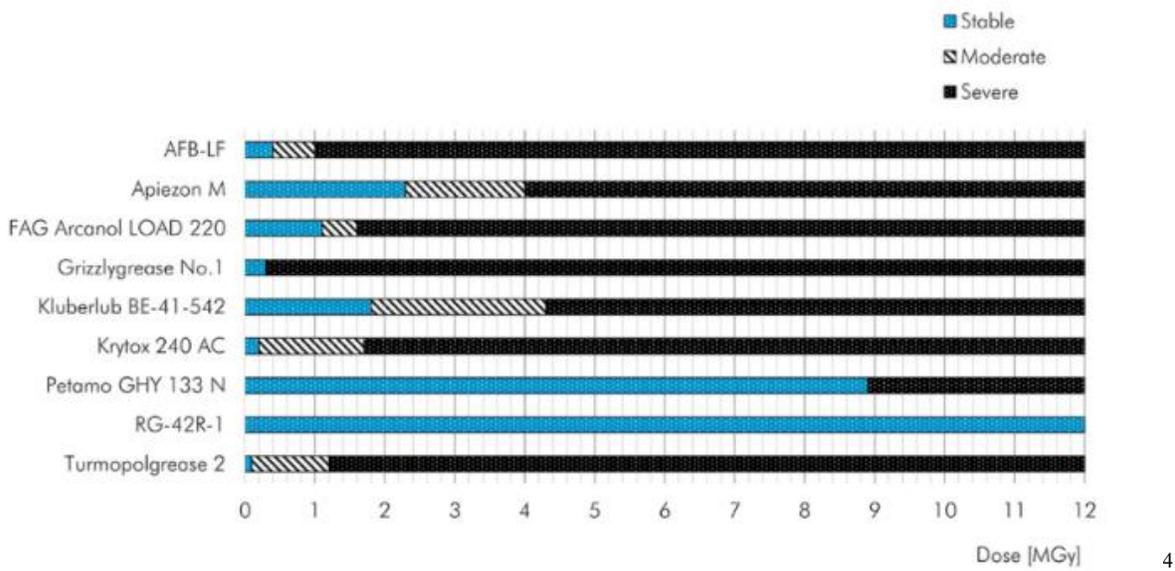

*Figure 4 - Damage thresholds of 9 commercial greases involved in study [36] and tested by standard worked penetration cone testing. The damage is defined by either 'Stable', Moderate', or 'Severe' based on variations of worked penetration testing. 'Stable' and 'moderate' being less than a 10% variation in the consistency and 'severe' indicating overall structural failure by solidification or liquefaction. [37]*

As for much of material science, the characterization of radiation effects on lubricants is dependent on the overall composition. This is an experimental endeavor, and only broad conclusions can be derived from any particular element or structure within the composition.

## 9. Radiation Effects on Bearings

Fortunately, little issue is found with the bearings and bearing housings themselves. Steel bearing assemblies are naturally hardened to radiation but have the disadvantage of needing lubricants which have proven to be a place of failure in ionizing radiation environments. Bearings that are made of ceramic do not use lubricants and therefore circumvent the issue of lubricants all together. However, ceramics undergo dramatic changes when irradiated and may lead to failures of their own. Radiation effects in ceramics are greatly dependent on the type of ceramic and certain ceramics are not suitable for being used in bearing assemblies. Some materials, such as zirconia, have very useful radiation resistant properties – undergoing damage due to slow moving ions but are repaired by swift ions. Because of this, zirconia may be of particular interest for motor bearing assemblies. [39] Carbon-based bearing may also be a material of interest in bearing requiring no lubrication. [40]

---

[4] Reprinted from Nuclear Materials and Energy, Vol 29, Matteo Ferrari, Dominika Senajova, Keith Kershaw, Antonio Perillo Marcone, Marco Calvian, 'Selection of radiation tolerant commercial greases for high-radiation areas at CERN: Methodology and applications', (2021), (https://doi.org/10.1016/j.nme.2021.101088). Creative Commons CC-BY.

## 10. Conclusion

There is much to consider when developing remote systems for radiation environments. Within complex systems, many different parts can be affected by radiation. As was reviewed in this paper, the stepper motor can be an effective tool for remote systems. The stepper motor is an electromagnetic motor that turns in discrete steps. In general, the stepper motor is quite radiation resistant, but several parts can have significant degradation. The first most radiation sensitive parts include those used in the controllers and drivers to operate the motor. The electronics involved may need to be contained in a radiation free area where the motor connection is maintained through long wires. The wires of the stepper motor an the next important consideration of stepper motor use in radiation environments. Irradiation will break down the insulation of the wire causing general impedance in the wire or electrical shorts. The permanent magnet that operates in conjunction with the wire coils can also see a decrease in magnetic strength after irradiation. Lastly, the lubricants involved in the ball bearings can also see significant degradation over time, causing increased friction and motor deterioration. Depending on the particular use of the stepper motor, alterations could be necessary for longer-term use. For instance, ceramic bearings could provide a lubricant free ball bearing option over the standard steel bearing with lubricant. For the purposes of this review, we will recommend materials that could improve stepper motor lifetime when used in radiation environments. First, the use of radiation resistant lubricants will improve lubricant lifetime. Two lubricants that have shown good radiation thresholds are PETAMO GHY 133 *N* and MORESCO RG-42R-1. For the permanent magnet, the use of a NdFeB magnets has shown to be a better choice than FeCrCo magnets. Pairing the NdFeB magnet with a more radiation resistant magnet like SmCo could improve magnet lifetime. Because of the extreme sensitivity of the electronics residing in the controller and driver, it is recommended these parts be removed from the radiation environment entirely and preserve connection through long wires. An alternative could also be the implementation of shielding around these components. Currently there are many shorter studies on radiation effects of stepper motors, but longer-term studies are lacking. Future long-term studies could provide important to determine the radiation thresholds of stepper motors and their parts. Further research may also prove important in identifying the most effective alterations in stepper motor parts for overall survivability in harsh radiation environments.

Acknowledgements: This work was funded by a grant from DOE, DE-SC0024815.

Ethics Statement: N/A

Data Availability Statement: N/A

Conflicts of Interest: Authors have no conflicts of interest to report.

# References

[1] M. J. Feldman and W. R. Hamel, "THE ADVANCEMENT OF REMOTE SYSTEMS TECHNOLOGY: PAST PERSPECTIVES AND FUTURE PLANS," Oak Ridge, TN, Oct. 1984.

[2] R. Sharp and M. Decreton, "Radiation tolerance of components and materials in nuclear robot applications," 1996.

[3] R. P. Ruilope, "Modelling and Control of Stepper Motors for High Accuracy Positioning Systems Used in Radioactive Environments," 2014.

[4] S. Murugesan, "An Overview of Electric Motors for Space Applications," 1981.

[5] A. J. Samin, "A review of radiation-induced demagnetization of permanent magnets," May 01, 2018, *Elsevier B.V.* doi: 10.1016/j.jnucmat.2018.02.029.

[6] B. W. Hotten and J. G. Carroll, "Radiation Damage in Lubricating Greases," 1957. [Online]. Available: https://pubs.acs.org/sharingguidelines

[7] K. Nordlund *et al.*, "Primary radiation damage: A review of current understanding and models," Dec. 15, 2018, *Elsevier B.V.* doi: 10.1016/j.jnucmat.2018.10.027.

[8] H. Shulman and W. S. Ginell, "NUCLEAR AND SPACE RADIATION EFFECTS ON MATERIALS," Jun. 1970.

[9] G. Von White II, R. Tandon, L. Serna, M. Celina, and R. Bernstein, "An Overview of Basic Radiation Effects on Polymers," Sep. 2013.

[10] H. T. Ngoc, T. T. Nguyen, H. Q. Cao, K. M. Le, and J. W. Jeon, "Improving the accuracy of permanent magnet rotor position estimation for stepper motors using magnetic induction and harmonic rejection," *IET Power Electronics*, vol. 13, no. 11, pp. 2236–2244, Aug. 2020, doi: 10.1049/iet-pel.2019.1128.

[11] Moons', "Step Motor – Basic Structure & Operation." Accessed: Jun. 23, 2025. [Online]. Available: https://www.moonsindustries.com/article/basic-structure-and-operating-principle-of-stepper-motor

[12] A. Suppan, M. Englert, and A. Rahn, "Irradiation of Motors for ln-Vessel Handling Equipment," 1997.

[13] S. Li, "On the Study of Radiation Sensitivity of Robot Components: Harmonic Drive and BLDC Motor - Dissertation," 2016.

[14] J. E. Ayer and G. J. Pokorny, "THE PERFORMANCE OF A MOTOR, A SWITCH, AND TWO TYPES OF PRESSURE PICKUP IN A HIGH-GAMMA-FLUX ENVIRONMENT," 1961.

[15] S. Li, A. Samin, Yuan. F. Zheng, and L. (Raymond) Cao, *The Effects of Radiaton-induced Demagnetization on the Performance of the Brushless DC Motor in Robot Servo Systems, 2014 International Symposium on Fundamentals of Electrical Engineering (ISFEE) : University Politehnica of Bucharest, Romania, November 28-29, 2014*. Bucharest, Romania: IEEE, 2014.

[16] M. Nakamichi, E. Ishitsuka, S. Shimakawa, and S. Kan, "Irradiation test of component for radiation-resistant small sized motor," *Fusion Engineering and Design*, vol. 84, no. 7–11, pp. 1399–1403, Jun. 2009, doi: 10.1016/j.fusengdes.2008.11.087.

[17] J. V. Draper, B. S. Weil, and J. B. Chesser, "RADIATION EFFECTS AND COMPONENT HARDENING TESTING PROGRAM AT THE OAK RIDGE NATIONAL LABORATORY," Oak Ridge, Tennessee, Apr. 1993.

[18] R. Lai *et al.*, *A Radiation-hard Gate Driver Circuit for High Voltage Application*. IEEE, 2020.

[19] K. E. Martin, M. K. Gauthier, J. R. Coss, A. R. V. Dantan, and W. E. Price, "Total-Dose Radiation Effects Data for Semiconductor Devices," Pasadena, California, 1985.

[20] D. K. Myers, "Ionizing Radiation Effects on Various Commercial NMOS Microprocessors," *IEEE Trans Nucl Sci*, vol. Vol. NS-24. No. 6, 1977.

[21] A. R. Hefnertt, D. L. Blackburnt, and K. F. Gallowaytt, "THE EFFECT OF NEUTRONS ON THE CHARACTERISTICS OF THE INSULATED GATE BIPOLAR TRANSISTOR (IGBT)*," 1986.

[22] R. P. Ruilope, "Modelling and Control of Stepper Motors for High Accuracy Positioning Systems Used in Radioactive Environments," 2014.

[23] M. C. Arva, D. N. M. Stanica, N. Bizon, and C. Ivan, "Statistical and sensitivity analysis of stepper motor parameters used in high gamma radiation field," in *Proceedings of the 13th International Conference on Electronics, Computers and Artificial Intelligence, ECAI 2021*, Institute of Electrical and Electronics Engineers Inc., Jul. 2021. doi: 10.1109/ECAI52376.2021.9515038.

[24] Y. Onishi *et al.*, "Study on polymer materials for development of the super 100 MGy-radiation resistant motor," *Polym J*, vol. 36, no. 8, pp. 617–622, 2004, doi: 10.1295/polymj.36.617.

[25] Z. Dong, Z. Mei, and K. Fan, "Optimal Design of Shielding for a Stepper Motor in an Energy Degrader System," in *2021 4th International Conference on Energy, Electrical and Power Engineering, CEEPE 2021*, Institute of Electrical and Electronics Engineers Inc., Apr. 2021, pp. 207–211. doi: 10.1109/CEEPE51765.2021.9475639.

[26] J. R. Cost, R. D. Brown, A. L. Giorgi, and J. T. Stanley, "Effects of Neutron Irradiation on Nd-Fe-B Magnetic Properties," 1988.

[27] R. Klaffky, R. Lindstrom, B. Maranville, R. Shull, B. J. Michlich, and J. Vacca, "THERMAL NEUTRON DEMAGNETIZATION OF NdFeB MAGNETS," Edinburgh, Scotland, 2006. [Online]. Available: www.srim.org

[28] L. R. Cao, A. Samin, and M. Kurth, "An analysis of radiation effects on NdFeB permanent magnets," *Nucl Instrum Methods Phys Res B*, vol. 342, pp. 200–205, Jan. 2015, doi: 10.1016/j.nimb.2014.10.006.

[29] T. Bizen *et al.*, "High-energy electron irradiation of NdFeB permanent magnets: Dependence of radiation damage on the electron energy," *Nucl Instrum Methods Phys Res A*, vol. 574, no. 3, pp. 401–406, May 2007, doi: 10.1016/j.nima.2007.01.185.

[30] S. Okamoto, T. Onishi, Y. Ueda, and T. Tanaka, "Acceleration Effects of Simultaneous Exposure to Heat and Ionizing Radiation on Deterioration of Insulating and Jacketing Materials of Electric Wires," *Bulletin of University of Osaka Prefecture*, no. Vol. 42, No. 2, pp. 185–192, Oct. 1993.

[31] C. G. Currin, "THE EFFECTS OF RADIATION ON SILICONE RUBBER WIRE AND CABLE INSULATION," Midland, Michagan, 1958.

[32] F. D. Terry, R. L. Kindred, and S. D. Anderson, "TRANSIENT NUCLEAR RADIATION EFFECTS ON TRANSDUCER DEVICES AND ELECTRICAL CABLES," Nov. 1965.

[33] F. Garner, K. Anderson, and T. Shikama, "RADIATION-INDUCED CHANGES IN ELECTRICAL CONDUCTIVITY OF A WIDE RANGE OF COPPER ALLOYS," *Fusion Reactor Materials Semiannual Progress Report DOE/ER-0313/9*, Jan. 1991.

[34] H. Schonbacher and M. Tavlet, "Compilation of Radiation Damage Test Data Part 1, 2nd Edition," Geneva, 1989.

[35] S. M. Hsu, "Molecular basis of lubrication," *Tribol Int*, vol. 37, no. 7, pp. 553–559, Jul. 2004, doi: 10.1016/j.triboint.2003.12.004.

[36] M. Ferrari *et al.*, "Experimental study of consistency degradation of different greases in mixed neutron and gamma radiation," *Heliyon*, vol. 5, no. 9, Sep. 2019, doi: 10.1016/j.heliyon.2019.e02489.

[37]     M. Ferrari, D. Senajova, K. Kershaw, A. Perillo Marcone, and M. Calviani, "Selection of radiation tolerant commercial greases for high-radiation areas at CERN: Methodology and applications," *Nuclear Materials and Energy*, vol. 29, Dec. 2021, doi: 10.1016/j.nme.2021.101088.

[38]     M. Ferrari, A. Zenoni, Y. Lee, and A. Andrighetto, "Radiation Effects in Greases for Use in Facilities Producing Intense Neutron Fields," Physical Society of Japan, Feb. 2020. doi: 10.7566/jpscp.28.061005.

[39]     L. Thomé, S. Moll, A. Debelle, F. Garrido, G. Sattonnay, and J. Jagielski, "Radiation effects in nuclear ceramics," 2012. doi: 10.1155/2012/905474.

[40]     R. K. Sreenilayam-Raveendran *et al.*, "Comparative evaluation of metal and polymer ball bearings," *Wear*, vol. 302, no. 1–2, pp. 1499–1505, Apr. 2013, doi: 10.1016/j.wear.2013.01.057.